\documentclass[aps,pra, twocolumn, amsmath, amssymb, showpacs, superscriptaddress]{revtex4}

\usepackage{graphicx}
\usepackage{bm}

\usepackage{xcolor}

\begin{document}

\title{Atomic interferometry based on a phase-jump error signal with suppression of the lineshape-asymmetry-caused shift}

\author{V.\,I.\,Yudin}
\email{viyudin@mail.ru}
\affiliation{Novosibirsk State University, ul. Pirogova 1, Novosibirsk, 630090, Russia}
\affiliation{Institute of Laser Physics SB RAS, pr. Akademika Lavrent'eva 15 B, Novosibirsk, 630090, Russia}
\affiliation{Novosibirsk State Technical University, pr. Karla Marksa 20, Novosibirsk, 630073, Russia}
\author{O.\,N.\,Prudnikov}
\affiliation{Novosibirsk State University, ul. Pirogova 1, Novosibirsk, 630090, Russia}
\affiliation{Institute of Laser Physics SB RAS, pr. Akademika Lavrent'eva 15 B, Novosibirsk, 630090, Russia}
\author{A.\,V.\,Taichenachev}
\affiliation{Novosibirsk State University, ul. Pirogova 1, Novosibirsk, 630090, Russia}
\affiliation{Institute of Laser Physics SB RAS, pr. Akademika Lavrent'eva 15 B, Novosibirsk, 630090, Russia}
\author{M.\,Yu.\,Basalaev}
\affiliation{Novosibirsk State University, ul. Pirogova 1, Novosibirsk, 630090, Russia}
\affiliation{Institute of Laser Physics SB RAS, pr. Akademika Lavrent'eva 15 B, Novosibirsk, 630090, Russia}
\affiliation{Novosibirsk State Technical University, pr. Karla Marksa 20, Novosibirsk, 630073, Russia}
\author{M.\,D.\,Radchenko}
\affiliation{Novosibirsk State University, ul. Pirogova 1, Novosibirsk, 630090, Russia}
\affiliation{Novosibirsk State Technical University, pr. Karla Marksa 20, Novosibirsk, 630073, Russia}
\author{D.\,N.\,Kapusta}
\affiliation{Institute of Laser Physics SB RAS, pr. Akademika Lavrent'eva 15 B, Novosibirsk, 630090, Russia}
\author{A.\,N.\,Goncharov}
\affiliation{Institute of Laser Physics SB RAS, pr. Akademika Lavrent'eva 15 B, Novosibirsk, 630090, Russia}
\author{A.\,E.\,Bonert}
\affiliation{Institute of Laser Physics SB RAS, pr. Akademika Lavrent'eva 15 B, Novosibirsk, 630090, Russia}
\author{V.\,G.\,Pal'chikov}
\affiliation{All-Russian Research Institute of Physical and Radio Engineering Measurements, Mendeleevo, Moscow region, 141570 Russia}
\affiliation{National Research Nuclear University MEPhI (Moscow Engineering Physics Institute), Moscow, 115409 Russia}
\author{T.~Zanon-Willette}
\affiliation{Sorbonne Universit$\acute{e}$, F-75005, Paris, France}
\author{L.\,Zhou}
\affiliation{Innovation Academy for Precision Measurement Science and Technology, Chinese Academy of Sciences, Wuhan 430071, China}
\author{M.\,S.\,Zhan}
\affiliation{Innovation Academy for Precision Measurement Science and Technology, Chinese Academy of Sciences, Wuhan 430071, China}


\begin{abstract}
{We develop a method for forming Ramsey fringes in atomic interferometers as a difference error signal using the phase-jump technique for Ramsey pulses (phase-jump error signal). In this case, the measurement procedure can be implemented as stabilization (locking) of the frequency chirp rate at the central zero of the error signal. We show analytically that this approach allows suppressing the lineshape-asymmetry-caused shift (LACS). This shift was recently theoretically predicted and experimentally identified in the papers [V. I. Yudin et al., JETP Lett. \textbf{122}, 737 (2025); D. N. Kapusta et al., JETP Lett. \textbf{123}, 388 (2026)]. A characteristic feature of the LACS is an inverse cubic dependence  $\propto 1/T^3$ (where $T$ is the time between adjacent Ramsey pulses), and it can become dominant for interferometers with a short baseline. Therefore, the developed measurement algorithm with a phase-jump error signal will enable significant improvement in the metrological performance of compact quantum sensors for various purposes (atomic gravimeters, accelerometers, gyroscopes).}
\end{abstract}

\maketitle

Modern atom interferometry is a fundamental platform for developing a new generation of high-precision quantum sensors. After the first demonstration of a quantum gravimeter based on cold atom interference \cite{Kasevich_1991}, this field has experienced rapid development, leading to the creation of devices with unprecedented sensitivity and long-term stability \cite{Peters_2001,Freier_2016,Menoret_2018,Heine_2020}. A distinctive feature of such devices is the absence of the need for regular calibration, which makes them in highly demand both for fundamental research (testing the equivalence principle, measuring the gravitational constant) \cite{Wicht_2002,Rosi_2014,Duan_2016,Asenbaum_2020,Zhou_2022} and applied tasks, including geodesy, navigation, and monitoring of natural resources, etc. \cite{Kennedy_2016,Bidel_2018,Bidel_2020,Vovrosh_2023,Diament_2024}.

One of the key trends in the development of atom interferometers is their miniaturization and mobility. In this context, short-baseline (short-time-of-flight) interferometers offer a promising alternative. They are less susceptible to external perturbations, enable high sampling rates (up to tens of hertz), and open up broad possibilities for the creation of compact devices. As demonstrated in \cite{McGuinness_2012,Lee_2022,Narducci_2022}, even with a sequence duration on the order of a few milliseconds, it is possible to achieve metrological characteristics sufficient for most practical applications. This is facilitated by new methods for enhancing phase stability and interference signal contrast \cite{Olivares_2020,Qing_2023,Wang_2024}, as well as by efficient data processing algorithms \cite{Cheiney_2018,Wei_2025}.

However, reducing the atomic transit time also gives rise to new systematic effects that previously had no significant impact. In our recent paper \cite{Yudin_2025}, we theoretically predicted a previously unknown lineshape-asymmetry-caused shift (LACS). A characteristic feature of this shift is its dependence $\propto 1/T^3$ on the time interval between pulses $T$, making it dominant specifically in short-baseline interferometers and significantly limiting the accuracy of compact quantum sensors. This effect was experimentally identified in \cite{Kapusta_2026}.

The present work is devoted to further theoretical analysis of the LACS shift in Mach-Zehnder atomic interferometers, as well as to the development of an effective method for its suppression. For this purpose,we compare two techniques for forming the difference error signal. In addition to the standard approach based on forming the error signal using the chirp-jump technique \cite{Santos_2011}, we consider an alternative scheme using control of the phases of individual Ramsey pulses (phase-jump technique). It is shown analytically that the standard chirp-jump error signal method does not eliminate the LACS shift, whereas the use of a phase-jump error signal allows completely suppressing this shift, including when averaging over an arbitrary velocity distribution of atoms.

Below we will follow the mathematical formalism presented in \cite{Yudin_2025}. As a specific scheme of an atomic interferometer, we consider a quantum gravimeter implemented on an atomic transition with unperturbed frequency $\omega_0$, where $|g\rangle$ and $|e\rangle$ are the ground and excited states, respectively, to which the following column vectors correspond:
\begin{equation}\label{basis}
|g\rangle=\left(
            \begin{array}{c}
              0 \\
              1 \\
            \end{array}
          \right),\quad
|e\rangle=\left(
            \begin{array}{c}
              1 \\
              0 \\
            \end{array}
             \right).
\end{equation}
In this two-level atom, the transition $|g\rangle$$\leftrightarrow$$|e\rangle$ is induced under the action of a resonant interaction operator of the type of a plane traveling wave with wave vector $\textbf{k}_{\rm eff}$
\begin{equation}\label{wave}
\widehat{V}(t,\textbf{r})=-\frac{\hbar\Omega_0}{2} e^{-i[\omega t+\alpha t^2/2-(\textbf{r}\cdot\textbf{k}_{\rm eff})+\phi]}|e\rangle\langle g|+h.c.\,,
\end{equation}
where $\omega$ is the carrier frequency, which can be varied near the transition frequency $\omega_0$; $\alpha$ is the rate of a frequency chirp; $\phi$ is the additional phase, which we can control; and $\Omega_0>0$ is the real Rabi frequency of the induced transitions. As is known, in such an interferometric scheme, the quantity $ (\textbf{g}\cdot \textbf{k}_{\rm eff})$ is measured, where $\textbf{g}$ is the gravitational acceleration.

Note that the type of interaction (\ref{wave}) takes place both for one-photon transitions under the action of a resonant traveling light wave in the case of an optical transition $|g\rangle$$\leftrightarrow$$|e\rangle$, and for two-photon transitions in a bichromatic field of two counter-propagating traveling light waves, when the difference of the optical frequencies of these waves ($\omega_1-\omega_2$) varies near the microwave frequency $\omega_0$ in the case of a transition  between hyperfine levels in the ground state (for example, in alkali metal atoms). However, this does not matter for our consideration.

\begin{figure}[t]
    \includegraphics[width=0.9\linewidth]{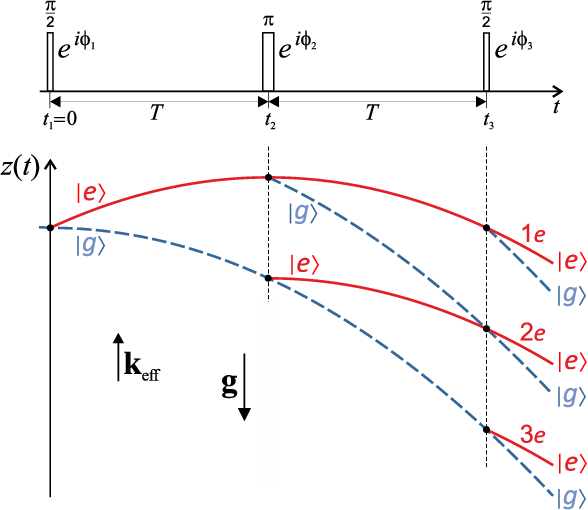}
    \caption{Standard scheme of a Mach-Zehnder type atomic interferometer-gravimeter.}\label{Rams_pulses}
\end{figure}

Let us consider a Mach-Zehnder atomic interferometer-gravimeter based on a sequence of three Ramsey pulses, shown in Fig.\,\ref{Rams_pulses}. In the reference frame of an atom moving along a classical trajectory, operator (\ref{wave}) can be formally represented as
\begin{equation}\label{atomic_ref}
\widehat{V}^{\rm (atom)}=-\frac{\hbar\Omega_0}{2} e^{-i\varphi(t)}|e\rangle\langle g|+h.c.\,,
\end{equation}
with a time-dependent phase, $\varphi(t)$. Then, for sufficiently short durations of the Ramsey pulses, the evolution operator of the atom in the basis (\ref{basis}) under the action of the $j$-th Ramsey pulse can be written as
\begin{equation}\label{j_evolution}
\widehat{R}^{(j)}=\left(
    \begin{array}{cc}
      A^{}_j & iB_j e^{-i\varphi(t_j)} \\
      iB_j e^{i\varphi(t_j)} & A^{\ast}_j \\
    \end{array}
  \right),
\end{equation}
where $t_j$ is the time of action of the $j$-th Ramsey pulse (see Fig.\,\ref{Rams_pulses}). The matrix elements in (\ref{j_evolution}) are determined as follows
\begin{eqnarray}\label{matrix_el}
&&A_j= \cos\bigg(\frac{\Omega^{(j)}\tau_j}{2}\bigg)+i\frac{\delta_j}{\Omega^{(j)}}\sin\bigg(\frac{\Omega^{(j)}\tau_j}{2}\bigg),\nonumber\\
&&B_j=\frac{\Omega_0}{\Omega^{(j)}}\sin\bigg(\frac{\Omega^{(j)}\tau_j}{2}\bigg),\\
&& \Omega^{(j)}=\sqrt{\Omega^2_0+\delta^2_j}\,,\nonumber
\end{eqnarray}
where $\tau_j$ is the duration of the $j$-th pulse, $\delta_j$ is the effective detuning of perturbation (\ref{atomic_ref}) from resonance during the action of the $j$-th pulse from the point of view of the moving atom, $\Omega^{(j)}$ is the generalized Rabi frequency during the action of the $j$-th pulse. Note that $B_j$ is a real number.

Let us consider the usual scheme of atom interrogation in the case of a spatially homogeneous gravitational acceleration $\textbf{g}$, when at time $t=0$ (i.e., at the beginning of the action of the 1st Ramsey pulse, see Fig.\,\ref{Rams_pulses}) the atom are in the lower state $|g\rangle$ and have an initial velocity $\textbf{v}$. Then, as the spectroscopic signal after the action of the Ramsey sequence in Fig.\,\ref{Rams_pulses}, we will consider the total population $N_e$ of atoms in the upper level $|e\rangle$ on three different trajectories: 1$e$, 2$e$ and 3$e$ (see red solid lines in Fig.\,\ref{Rams_pulses}). Assuming that after the action of three Ramsey pulses all three trajectories have diverged to distances noticeably larger than the de Broglie wavelength, we have the following expression
\begin{equation}\label{N_e}
N_e=n_{1e}+n_{2e}+n_{3e}=|K_{1e}|^2+|K_{2e}|^2+|K_{3e}|^2.
\end{equation}
Using the notation from (\ref{j_evolution}), we can write
\begin{eqnarray}\label{K}
  K_{1e} &=&B_1 A_2 A_3,\nonumber  \\
  K_{2e} &=&B_1 B_2 B_3- A^{\ast}_1 B_2 A_3 e^{i\widetilde{\alpha}T^2}e^{i\Phi}, \\
  K_{3e} &=& A^{\ast}_1 A^{\ast}_2 B_3,\nonumber
\end{eqnarray}
where $T$ is the free evolution time between pulses, and $\widetilde{\alpha}$ is the frequency chirp rate centered on the measured quantity $ (\textbf{g}\cdot \textbf{k}_{\rm eff})$
\begin{equation}\label{alpha}
\widetilde{\alpha}=\alpha - (\textbf{g}\cdot \textbf{k}_{\rm eff}).
\end{equation}
In the standard interrogation scheme, we have the following expressions for the duration and one-photon detuning of each pulse
\begin{eqnarray}\label{delta_j}
&& \tau_1=\tau_3=\tau,\quad \tau_2=2\tau, \nonumber\\
&& \delta_1=\Delta_{\rm D},\quad \delta_2= \widetilde{\alpha}T+\Delta_{\rm D},\quad  \delta_3= 2\widetilde{\alpha}T+\Delta_{\rm D},
\end{eqnarray}
where $\Delta_{\rm D}$ is the residual Doppler detuning from resonance during the action of each pulse, which, taking into account the recoil effect for atoms with mass $M$, has the following standard form
\begin{equation}\label{Delta}
\Delta_{\rm D}=\omega-\omega_0-\frac{\hbar|\textbf{k}_{\rm eff}|^2}{2M}-(\textbf{v}\cdot \textbf{k}_{\rm eff})-\overline{\delta}_{\rm ac}\,,
\end{equation}
where $(\textbf{v}\cdot \textbf{k}_{\rm eff})$ is the Doppler frequency shift during the action of the 1st Ramsey pulse at $t=0$ (see Fig.\,\ref{Rams_pulses}). In (\ref{Delta}), we also took into account the Stark shift $\overline{\delta}_{\rm ac}$ of the atomic transition frequency caused by the probe light field, and which we assume to be the same for all three pulses.

In addition, in contrast to \cite{Yudin_2025}, for our further purposes we introduced additional (artificial) phases of the pulses $\phi_1$, $\phi_2$ and $\phi_3$  (see Fig.\,\ref{Rams_pulses}), which we can control individually for each Ramsey pulse. These phases determine the parameter $\Phi$:
\begin{equation}\label{Phi}
\Phi=\phi_1+\phi_3-2\phi_2\,,
\end{equation}
which is contained in the additional phase factor $e^{i\Phi}$ in the expression for $K_{2e}$ in (\ref{K}).

The functional dependence of the spectroscopic signal $N_e(\widetilde{\alpha},\Phi)$ on the varied parameter $\widetilde{\alpha}$ at the fixed $\Phi$ can be represented as
\begin{equation}\label{N_Rams}
N_e(\widetilde{\alpha},\Phi)=|K_{1e}(\widetilde{\alpha})|^2+|K_{2e}(\widetilde{\alpha},\Phi)|^2+|K_{3e}(\widetilde{\alpha})|^2,
\end{equation}
where we took into account the phase insensitivity (with respect to phase $\Phi$) of the two contributions $|K_{1e}(\widetilde{\alpha})|^2$ and $|K_{3e}(\widetilde{\alpha})|^2$. In the general case, the spectral dependence $N_e(\widetilde{\alpha},\Phi)$ on the variable $\widetilde{\alpha}$ represents a sequence of narrow Ramsey resonances with FWHM about $\pi/T^2$ with a flat envelope (width of the order of $\sim 2\Omega_0/T$), which is described by the interference contribution $|K_{2e}(\widetilde{\alpha},\Phi)|^2$ in (\ref{N_Rams}). In addition, there is also a broad substrate with a flat dependence (width of the order of $\sim 2\Omega_0/T$), to which the other two terms $|K_{1e}(\widetilde{\alpha})|^2$ and $|K_{3e}(\widetilde{\alpha})|^2$ contribute.

In [24], we considered the usual case of equal phases of all three pulses $\phi_1$=$\,\phi_2$=$\,\phi_3$ (i.e., $\Phi=0$), when the gravimetric measurement of the quantity $(\textbf{g}\cdot \textbf{k}_{\rm eff})$ corresponds to the position of the top of the central Ramsey resonance in the dependence $N_e(\widetilde{\alpha},0)$. As a result, in [24], the lineshape-asymmetry-caused shift (LACS) was found in the case of $\Delta_{\rm D}\neq 0$. It was shown that this shift is proportional to $1/T^3$ and it can become dominant for short-baseline interferometers.

\begin{figure}[t]
    \includegraphics[width=0.8\linewidth]{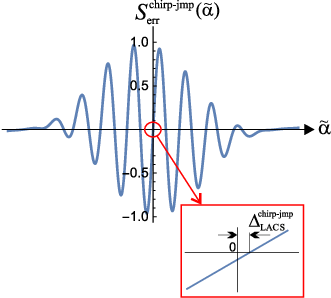}
    \caption{Demonstration of the shift in the position of the central zero for the chirp-jump error signal (for $\Delta_{\rm D}\neq 0$).}\label{ch_jmp_Err}
\end{figure}

However, in addition to gravimetric measurements based on the position of the top of the central Ramsey resonance, another algorithm proposed in \cite{Santos_2011} is also used. As applied to our formulation of the problem, the essence of the approach \cite{Santos_2011} can be expressed as the formation of an error signal $S^\text{chirp-jmp}_\text{err}(\widetilde{\alpha})$ by subtracting measurements at two points $\widetilde{\alpha}\pm\Delta\alpha$ (i.e., using the chirp-jump technique):
\begin{equation}\label{S_ch_jmp}
S^\text{chirp-jmp}_\text{err}(\widetilde{\alpha})=N_e(\widetilde{\alpha}+\Delta\alpha,0)-N_e(\widetilde{\alpha}-\Delta\alpha,0)\,,
\end{equation}
where $\Delta\alpha$ is some fixed value of the chirp-rate jump. We will call such a difference signal as a chirp-jump error signal. The dispersive lineshape of the error signal $S^\text{chirp-jmp}_\text{err}(\widetilde{\alpha})$ is presented in Fig.\,\ref{ch_jmp_Err}. In this case, the gravimetric measurement is determined by the position of the central zero, which is found from the equation $S^\text{chirp-jmp}_\text{err}(\widetilde{\alpha})=0$. To maximize the slope at the center of the dispersion curve, one should use the value $\Delta\alpha=0.5\pi/T^2$ for the chirp-rate jump.

\begin{figure}[t]
    \includegraphics[width=0.85\linewidth]{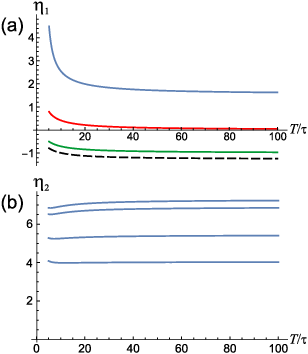}
    \caption{(a)\;Dependence $\eta^{}_1(T/\tau,T^2\Delta\alpha)$ on the dimensionless parameter $T/\tau$ in the expression (\ref{LACS_shirp}) for several fixed values of the chirp jump $T^2\Delta\alpha$=0,\,$\pi$/4,\,$\pi$/2,\,2$\pi$/3 (from bottom to top) in the case of $\Omega_0\tau$=$\pi/2$. (b)\;Dependence of the dimensionless function $\eta^{}_2(T/\tau,|{\bf k}_{\rm eff}|\bar{v}/\Omega_0)$ on $T/\tau$ in the expression (\ref{LACS_shirp_v}) for several fixed values $|{\bf k}_{\rm eff}|\bar{v}/\Omega_0\,$=\,5,\,10,\,40,\,100 (from bottom to top) for $\Omega_0\tau=\pi/2$ and $\Delta\alpha=0.5\pi/T^2$.}\label{fT_ch_jmp}
\end{figure}

Using the symmetry analysis developed in \cite{Yudin_2025}, it is not difficult to show that a shift also appears in the gravimetric measurement using $S^\text{chirp-jmp}_\text{err}(\widetilde{\alpha})$ in the presence of the residual Doppler shift $\Delta_{\rm D}\neq 0$. Indeed, let us consider the error signal (\ref{S_ch_jmp}) at the point $\widetilde{\alpha}=0$:
\begin{equation}\label{ch_jmp_zerr}
S^\text{chirp-jmp}_\text{err}(0)=N_e(\Delta\alpha,0)-N_e(-\Delta\alpha,0).
\end{equation}
Then, as follows from \cite{Yudin_2025}, where the asymmetry of the resonance $N_e(\widetilde{\alpha},0)\neq N_e(-\widetilde{\alpha},0)$ was shown for $\Delta_{\rm D}\neq 0$, we obtain $S^\text{chirp-jmp}_\text{err}(0)\neq 0$ for $\Delta_{\rm D}\neq 0$. Consequently, the zero of the error signal is located at the point $\widetilde{\alpha}=\Delta^\text{chirp-jmp}_{\rm LACS}\neq 0$ (see Fig.\,\ref{ch_jmp_Err}). In this case, the shift $\Delta^\text{chirp-jmp}_{\rm LACS}$ also depends on the value of $\Delta\alpha$ and may differ noticeably from the shift in \cite{Yudin_2025}. Numerical calculations show that for $\Omega_0\tau=\pi/2$ the LACS shift is described rather well by the expression
\begin{equation}\label{LACS_shirp}
\Delta^\text{chirp-jmp}_{\rm LACS}\approx \eta^{}_1(T/\tau,T^2\Delta\alpha)\frac{\tau}{T^3}\frac{\Delta_{\rm D}}{\Omega_0}\,,
\end{equation}
where the dimensionless function $\eta^{}_1(T/\tau,T^2\Delta\alpha)$ depends on two dimensionless parameters: $T/\tau$ and $T^2\Delta\alpha$. The forms of this function for several values of $T^2\Delta\alpha$ are presented in Fig.\,\ref{fT_ch_jmp}(a) in the region $T$$\gg$$\,\tau$. It is interesting to note that for a mono-velocity atomic ensemble, the magnitude and sign of the shift substantially depend on the magnitude of $\Delta\alpha$. At the same time, for the optimal value $\Delta\alpha=0.5\pi/T^2$, the LACS shift becomes relatively small [see the red line in Fig.\,\ref{fT_ch_jmp}(a)].

In the case of a non-mono-velocity atomic ensemble, instead of (\ref{S_ch_jmp}), one should use the expression for the error signal averaged over the velocity distribution $f(\bf v)$
\begin{align}\label{S_ch_jmp_v}
\langle S^\text{chirp-jmp}_\text{err}(\widetilde{\alpha})\rangle^{}_{\bf v}&=\\
&\langle N_e(\widetilde{\alpha}+\Delta\alpha,0)\rangle^{}_{\bf v}-\langle N_e(\widetilde{\alpha}-\Delta\alpha,0)\rangle^{}_{\bf v},\nonumber
\end{align}
where the symbol $\langle W\rangle^{}_{\bf v}$ means:
\begin{equation}\label{v}
\langle W\rangle^{}_{\bf v}=\int Wf({\bf v})d{\bf v}^3.
\end{equation}
Below in the calculations we will consider a Maxwellian velocity distribution
\begin{equation}\label{fv_M}
 f_{\rm M}({\bf v})=\frac{1}{\pi^{3/2}\bar{v}^3}e^{-\frac{({\bf v}-{\bf v}_0)^2}{\bar{v}^2}},
\end{equation}
where $\bar{v}$ is the characteristic width of the velocity distribution relative to the mean velocity of atoms ${\bf v}_0$.

In this case, the magnitude of the LACS undergoes substantial changes. For example, as numerical calculations show, for $\Omega_0\tau\,$=$\,\pi/2$ and a phase jump $\Delta\alpha=0.5\pi/T^2$ in the case when the Doppler width $|{\bf k}_{\rm eff}| \bar{v}$ substantially exceeds the Rabi frequency ($|{\bf k}_{\rm eff}| \bar{v}>5 \Omega_0$), the following expression works well
\begin{equation}\label{LACS_shirp_v}
\Delta^{\langle \text{chirp-jmp}\rangle_{\bf v}}_{\rm LACS}\approx \eta^{}_2(T/\tau,|{\bf k}_{\rm eff}| \bar{v}/\Omega_0) \frac{\tau}{T^3}\frac{\Omega_0}{|{\bf k}_{\rm eff}| \bar{v}}\frac{\langle\Delta_{\rm D}\rangle_{\bf v}}{|{\bf k}_{\rm eff}| \bar{v}}\,,
\end{equation}
where
\begin{equation}\label{Delta_v}
\langle\Delta_{\rm D}\rangle^{}_{\bf v}=\omega-\omega_0-\frac{\hbar|\textbf{k}_{\rm eff}|^2}{2M}-(\textbf{v}_0\cdot \textbf{k}_{\rm eff})-\overline{\delta}_{\rm ac}\,,
\end{equation}
is the Doppler detuning for atoms with the mean velocity in the ensemble ${\bf v}_0$. The dimensionless function $\eta^{}_2(T/\tau,|{\bf k}_{\rm eff}| \bar{v}/\Omega_0)$ in (\ref{LACS_shirp_v}) depends rather slowly on the two dimensionless parameters $T/\tau$ and $|{\bf k}_{\rm eff}|\bar{v}/\Omega_0$ [see Fig.\,\ref{fT_ch_jmp}(b)].

Thus, for the chirp-jump error signal, the LACS shift is also existing, which is proportional to $1/T^3$ (as well as in \cite{Yudin_2025}).Its metrological significance on the example of atomic interferometer-gravimeters was estimated in \cite{Yudin_2025,Kapusta_2026}.

In the present paper, to suppress the LACS shift, we propose an alternative method of forming the error signal. For this, we will artificially change the phase of one of the pulses (for example, the third) relative to the other two so that a phase parameter $\Phi\neq 0$ appears for $K_{2e}(\widetilde{\alpha},\Phi)$ in the expression (\ref{K}). Then we can form the following error signal based on the difference of two signals with opposite phases $\pm\Phi$:
\begin{align}\label{S_ph_jmp}
S^\text{phase-jmp}_\text{err}(\widetilde{\alpha})&=N_e(\widetilde{\alpha},\Phi)-N_e(\widetilde{\alpha},-\Phi) =\nonumber\\
& |K_{2e}(\widetilde{\alpha},\Phi)|^2-|K_{2e}(\widetilde{\alpha},-\Phi)|^2,
\end{align}
where, in accordance with expression (\ref{N_Rams}), we took into account that the terms $|K_{1e}(\widetilde{\alpha})|^2$ and $|K_{3e}(\widetilde{\alpha})|^2$ do not depend on the phase $\Phi$. The differential signal (\ref{S_ph_jmp}) we will call as the phase-jump error signal. In this case, the gravimetric measurement is also determined by the position of the central zero, which is determined from the equation $S^\text{phase-jmp}_\text{err}(\widetilde{\alpha})=0$.

\begin{figure}[t]
    \includegraphics[width=0.8\linewidth]{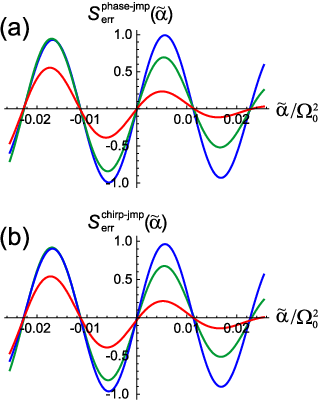}
    \caption{Error signals: (a)\;$S^\text{phase-jmp}_\text{err}(\widetilde{\alpha})$ for $\Phi=\pi/2$; (b)\;$S^\text{chirp-jmp}_\text{err}(\widetilde{\alpha})$ for $\Delta\alpha=0.5\pi/T^2$. Calculations were performed for various values of $\Delta_{\rm D}$: $\Delta_{\rm D}=0$ (blue line), $\Delta_{\rm D}/\Omega_0=0.5$ (green line),  $\Delta_{\rm D}/\Omega_0=1$ (red line). Calculation parameters: $\Omega_0\tau=\pi/2$, $T=10\tau$.}\label{PJ_err}
\end{figure}

Let us show that, in contrast to the chirp-jump error signal $S^\text{chirp-jmp}_\text{err}(\widetilde{\alpha})$, the phase-jump error signal $S^\text{phase-jmp}_\text{err}(\widetilde{\alpha})$ does not experience a shift even in the case of $\Delta_{\rm D}\neq 0$, i.e., $\Delta^\text{phase-jmp}_{\rm LACS}=0$. Indeed, let us consider the error signal at the point $\widetilde{\alpha}=0$:
\begin{equation}\label{S_ph_jmp0}
S^\text{phase-jmp}_\text{err}(0)= |K_{2e}(0,\Phi)|^2-|K_{2e}(0,-\Phi)|^2.
\end{equation}
Since, as follows from (\ref{delta_j}), at $\widetilde{\alpha}=0$ the first and third detunings are equal, $\delta_1=\delta_3$, then, in accordance with definition (\ref{matrix_el}), the equality for the matrix elements $A_1=A_3$ holds for an arbitrary value of $\Omega_0\tau$. As a result, based on (\ref{K}), we can write the following expressions
\begin{eqnarray}\label{K_ph_jmp}
&&K_{2e}(0,\Phi) =B_1 B_2 B_3- B_2 |A_3|^2 e^{i\Phi},\nonumber \\
&&K_{2e}(0,-\Phi) =B_1 B_2 B_3- B_2 |A_3|^2 e^{-i\Phi},
\end{eqnarray}
in which all coefficients are real, except for the phase factors $e^{\pm i\Phi}$. Thus, the equality $K^{}_{2e}(0,-\Phi)=K^{^*}_{2e}(0,\Phi)$ holds, which in expression (\ref{S_ph_jmp0}) automatically leads to the result
\begin{equation}\label{zerr_S_ph_jmp_2}
S^\text{phase-jmp}_\text{err}(0)=0,
\end{equation}
for any $\Delta_{\rm D}\neq 0$, $\Omega_0\tau$ and $\Phi$, i.e., the LACS is completely absent. For example, Fig.\,\ref{PJ_err}(a) shows the calculation of $S^\text{phase-jmp}_\text{err}(\widetilde{\alpha})$ for various values of the residual Doppler detuning $\Delta_{\rm D}$. For comparison, Fig.\,\ref{PJ_err}(b) shows the calculation of the chirp-jump error signal $S^\text{chirp-jmp}_\text{err}(\widetilde{\alpha})$ for the same values of $\Delta_{\rm D}$. As can be seen, the shapes of the signals $S^\text{phase-jmp}_\text{err}(\widetilde{\alpha})$ and $S^\text{chirp-jmp}_\text{err}(\widetilde{\alpha})$ are almost identical, except for the presence of the LACS shift in the center for the chirp-jump error signal in Fig.\,\ref{PJ_err}(b).

It should be noted that the phase-jump error signal (\ref{S_ph_jmp}) contains a contribution only from atoms on trajectory 2$e$ (see Fig.\,\ref{Rams_pulses}), on which the interference pattern is formed. While the chirp-jump error signal (\ref{S_ch_jmp}) contains also a contribution from the broad substrate (which is, in the general case, asymmetric in $\widetilde{\alpha}$, see \cite{Yudin_2025}), formed by atoms on trajectories 1$e$ and 3$e$ (see Fig.\,\ref{Rams_pulses}).

The independence of equality (\ref{zerr_S_ph_jmp_2}) from the values of $\Delta_{\rm D}$ means the independence from the magnitude of the Doppler shift $(\textbf{v}\cdot \textbf{k}_{\rm eff})$, which, in turn, means that the equality
\begin{equation}\label{zerr_S_ph_jmp_3}
 \langle S^\text{phase-jmp}_\text{err}(0)\rangle^{}_{\bf v}=0,
\end{equation}
also holds for the velocity-averaged error signal with any velocity distribution $f({\bf v})$ (i.e., not only for (\ref{fv_M})). Thus $\Delta^{\langle \text{phase-jmp}\rangle_{\bf v}}_{\rm LACS}=0$ for arbitrary $\langle\Delta_{\rm D}\rangle^{}_{\bf v}\neq 0$, $\Omega_0\tau$, and $\Phi$.

Note that the maximal slope of the error signal at the center of the line is achieved at $\Phi=\pi/2$. At the same time, from a theoretical point of view, it is completely unimportant how the formation of phase jumps $\pm\Phi$ is carried out. However, from a practical viewpoint, it seems most convenient to apply phase jumps $\phi_3=\pm\pi/2$ for the third pulse at $\phi_1=\phi_2=0$ (see Fig.\,\ref{Rams_pulses}).

Thus, the use of a phase-jump error signal allows suppressing the LACS regardless of the velocity distribution of atoms. This can significantly improve the metrological characteristics of small-sized atomic interferometers with a short baseline, without reducing the sensitivity of measurements.

As for other systematic shifts that appear in real experiments due to various deviations from the considered ideal interferometer scheme (spatial inhomogeneity of light and magnetic fields, imperfection of pulse shapes, etc.), the phase-jump and chirp-jump techniques are approximately identical with respect to them.

In conclusion, we have continued the theoretical study of the systematic lineshape-asymmetry-caused shift (LACS), predicted in \cite{Yudin_2025} for Mach-Zehnder atomic interferometers. Two approaches to the measurement procedure based on the formation of an error signal were analyzed. In the case of the standard chirp-jump technique, it is shown that the LACS is preserved and has the same dependence $\propto 1/T^3$.

The key result of the work is the proposal and justification of an alternative method for forming the error signal based on controlled phase jumps (phase-jump error signal). In contrast to the standard chirp-jump approach \cite{Santos_2011}, the proposed error signal is formed by subtracting signals obtained at two opposite values of the artificial phase $(\pm\Phi)$. It is analytically proven that for such an error signal, the position of the central zero does not depend on the Doppler detuning and, consequently, on the lineshape asymmetry. This also holds when averaging over an arbitrary velocity distribution of atoms. Thus, the use of a phase-jump error signal allows suppressing the LACS shift, that cannot be achieved using the standard chirp-jump error signal.

It is important to note that the proposed method does not require modification of the hardware part of the interferometer and can be implemented only at the level of the algorithm for controlling the phases of Ramsey pulses. In this case, the maximum slope of the dispersion curve of the error signal is achieved at $\Phi = \pi/2$, which ensures optimal measurement sensitivity.

The obtained results are of immediate significance for the development of small-sized atomic interferometers for various purposes (gravimeters, accelerometers, and gyroscopes). Suppression of the LACS shift can significantly improve their metrological characteristics, especially in conditions where the atomic transit time is limited by requirements of compactness and speed. The developed approach can also be extended to other configurations of atomic interferometers and coherent systems where the asymmetry of spectral lines plays a noticeable role.

Further research should focus on the experimental implementation of the proposed method, as well as on evaluating its robustness to other types of systematic errors and to the limiting parameters of the equipment used.

\end{document}